# Fourier phase-demodulation applied to strip-light 360-degrees profilometry of 3D solids; theoretical principles

**Manuel Servin, Moises Padilla and Guillermo Garnica**

*Centro de Investigaciones en Optica A. C., 115 Loma del Bosque,*
*Col. Lomas del Campestre, 37150 Leon Guanajuato, Mexico.*

**October 2015**

[*]mservin@cio.mx

**Abstract:** 360-degrees digitalization of three-dimensional (3D) solids using a projected light-strip is a well established technique. These profilometers project a light-strip over the solid under analysis while the solid is rotated a full revolution. Then a computer program typically extracts the centroid of this light-strip, and by triangulation one obtains the shape of the solid. Here instead of using intensity-based strip centroid estimation, we propose to use Fourier phase-demodulation. This 360-degrees profilometer first constructs a carrier-frequency fringe-pattern by closely adding individual light-strip images. Secondly this high-density fringe-pattern is phase-demodulated using the standard Fourier technique.

**OCIS codes:** (120.0120) Instrumentation, measurement, and metrology; (120.5050) Phase measurement; (120.4630) Optical inspection

**References and links**

1. M. Takeda, H. Ina, S. Kobayashi, "Fourier-transform method of fringe-pattern analysis for computer-based topography and interferometry," J Opt. Soc. Am. A **72**, l56–60 (1982).
2. M. Halioua, R. S. Krishnamurthy, H. C. Liu, and F.P. Chiang, "Automated 360$^o$ profilometry of 3-D diffuse objects," Appl. Opt. **24**, 2193-2196 (1985).
3. X. X. Cheng, X. Y. Su, and L. R. Guo, "Automated measurement method for 360$^o$ profilometry of 3-D diffuse objects," Appl. Opt., **30**, 1274-1278 (1991).
4. A. K. Asundi, "360-deg profilometry: new techniques for display and acquisition," Opt. Eng. **33**, 2760-2769 (1994).
5. M. Chang, W. C. Tai, "360-deg profile noncontact measurement using a neural network," Opt. Eng. **34**, 3572-3576 (1995).
6. A. S. Gomes, L. A. Serra, A. S. Lage, A. Gomes, "Automated 360 degree profilometry of human trunk for spinal deformity analysis," in Proceedings of Three Dimensional Analysis of Spinal Deformities, M. Damico et al. eds., (IOS, Burke, Virginia USA, 1995), pp. 423-429.
7. Y. Song, H. Zhao, W. Chen, and Y. Tan, "360 degree 3D profilometry," Proc. SPIE **3204**, 204-208 (1997).
8. A. Asundi and W. Zhou, "Mapping algorithm for 360-deg profilometry with time delayed integration imaging," Opt. Eng. **38**, 339-344 (1999).
9. X. Su and W. Chen, "Fourier transform profilometry: a review," Opt. and Lasers in Eng. **35**, 263–284 (2001).
10. X. Zhang, P. Sun, H. Wang, "A new 360 rotation profilometry and its application in engine design," Proc. SPIE **4537**, 265-268 (2002).
11. J. A. Munoz-Rodriguez, A. Asundi, R. Rodriguez-Vera, "Recognition of a light line pattern by Hu moments for 3-D reconstruction of a rotated object," Opt. & Laser technology **37**, 131-138 (2005).
12. G. Tmjillo-Schiaffino, N. Portillo-Amavisca, D. P. Salas-Peimbert, L. Molina-de la Rosa, S. Almazan-Cuellarand, L. F. Corral-Martinez, "Three-dimensional profilometry of solid objects in rotation," in AIP Proceedings **992**, 924-928 (2008).
13. B. Shi, B. Zhang, F. Liu, J. Luo, and J. Bai, "360$^o$ Fourier transform profilometry in surface reconstruction for fluorescence molecular tomography," IEEE Journal Biomed. Health Inf. **17**, 681-689 (2013).
14. Y. Zhang and G. Bu, "Automatic 360-deg profilometry of a 3D object using a shearing interferometer and virtual grating", Proc. SPIE **2899**, 162-169 (1996).
15. M. Servin, G. Garnica, J. C. Estrada, and J. M. Padilla, "High-resolution low-noise 360-degree digital solid reconstruction using phase-stepping profilometry," Opt. Exp. **22**, 10914-10922 (2014).

## 1. Introduction

Fringe projection profilometry of three dimensional (3D) solids using Fourier phase-demodulation is well known since the classical paper by Takeda et al. in 1982 [1]. Although this phase-measuring technique effectively demonstrated that 3D digitalization was possible using a single carrier-frequency fringe-pattern, it cannot digitize the 360 degrees (360$^o$) view of the 3D solid. As far as we know this was first implemented by Halioua et al. in 1985 [2]. He used a linear fringe-pattern projection with a 3-step phase shifter and a turntable to obtain the 360 degrees profilometry of a mannequin head [2]. Later in 1991, Cheng et al. achieved depth estimation by the use of an artificial-neural network projecting a laser strip over a solid lying in a turntable [3]. Single laser-strip profilometers use triangulation-based height estimation [3], which (in general) are less accurate than phase demodulation of high-density, carrier-frequency fringe-patterns [1-2]. Asundi published a fast 360 degrees technique based on a stroboscope strip-light projection and a digital drum camera rotating the solid at 2500 rpm [4]. Chang et al. used laser light-strip [5] and reconstructed a solid with 360-degrees using a neural network to estimate the center of the strips. Gomes et al. projected a linear-grating to analyze the spinal deformities of a human; they used Fourier profilometry [6]. Later on Song et al. used a fringe grating projector and phase-shifting interferometry to obtain the 360-degree object [7]. Asundi et al. also implemented a very fast 360 degrees profilometer using a time delay integration imaging for digitalization [8]. The state of the art on 3D profilometry was reviewed in 2001 by Sue and Chen but they have just included a single paper of 360-degrees profilometry [9]. More recently Zhang et al. have used 360 degrees profilometry for flow analysis in mechanical engines [10]. In 2005 Munoz-Rodriguez et al. used triangulation for 3D object reconstruction by projecting a strip-light and Hu moments [11]. In 2008 Trujillo-Shiaffino et al. used 360-degrees profilometry based on strip-light projection and triangulation to digitize a smooth rotation-symmetric object [12]. More recently Shi et al. used 360-degrees profilometry applied to fluorescent molecular tomography [13]. Some researchers have used shearing interferometry to project high quality linear fringes for 360-degrees profilometry [14]. This is a self-contained view of the main ideas behind 360$^o$ profilometry to this date.

Here we describe the theoretical aspects of a 360-degree profilometer using a projected light-strip over the 3D-solid under analysis. We take *N* CCD-images of the projected line over the 3D-solid for each rotating angle-step. In other words, for each rotation angle-step one takes one camera-image while the solid is rotated a full 360-degrees revolution. With these *N*-images set, a carrier-fringe pattern is constructed. This resulting carrier-frequency fringe-pattern is finally analyzed using the Fourier phase-demodulation technique.

## 2. 360$^o$ profilometers which uses a light-strip projection

Due to the light-strip projection geometry shown in Fig. 1 one minimizes the self-occluding shadows cast over the sphere shown. This is a great advantage of light-strip 360-degrree profilometry. In contrast, the fringe-projection 360$^o$ profilometry in Ref. [2] cast more self-occluding shadows over the digitizing solids. Probably this is the reason why the technique in [2] has lower popularity than light-strip 360$^o$ profilometry. In any 360$^o$ profilometer, the solid (the sphere in Fig. 1) must be positioned over a turntable and rotated 360$^o$ in the azimuthal angle $\varphi$ to obtain the 3D-surface data from all possible 360$^o$ perspectives.

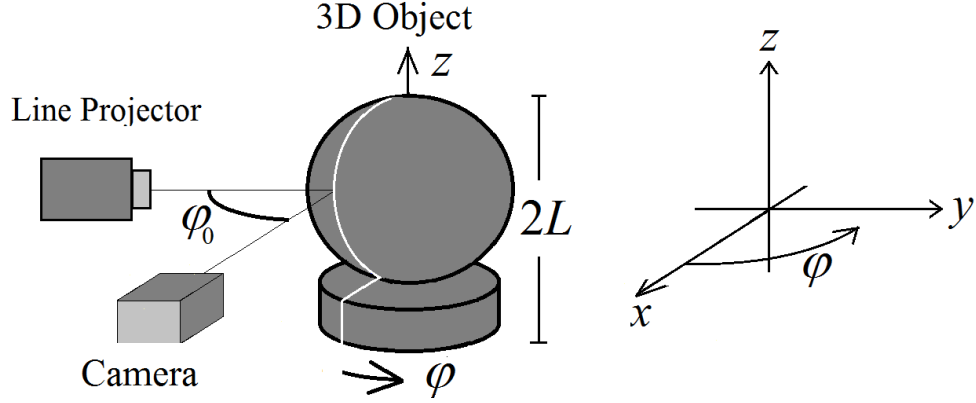

Fig. 1. This figure shows the light-strip profilometer with height-sensitivity $g=\sin(\varphi_0)$; with this set-up, the self-occluding shadows are minimized. From $N$-camera images a single carrier fringe-pattern is obtained which can be phase-demodulated using the Fourier technique. Please note that the viewing point of this figure it is not the camera's perspective.

A mathematical model for the intensity of a light-strip as imaged over the digitizing CCD camera plane $(y,z) \in \mathbb{R}^2$ may be the following Gaussian irradiance (Figs. 1-3),

$$G(y,z;\varphi) = \left\{ a(y,z;\varphi) + b(y,z;\varphi) e^{\frac{-[y-y(\varphi,z)]^2}{\sigma^2}} \right\} \prod(y); \qquad g=\sin(\theta_0), \varphi \in [0, 2\pi). \quad (1)$$

Where $y(\varphi,z)$ is the strip-light phase displacement shown in Fig. 2 and Fig. 3. The profilometry sensitivity is given by $g = \sin(\varphi_0)$. The window function $\prod(y)$ is an indicator relation (Fig. 2 and Fig. 3) which may be expressed as,

$$\prod(y) = \begin{cases} 1 & if \quad y \in [0, -\max[y(\varphi,z)]] \\ 0 & otherwise \end{cases} ; \quad \varphi \in [0, 2\pi], \quad z \in [-L.L]. \quad (2)$$

The camera takes the image of $G(y,z;\varphi)$ as Fig. 2 shows. The center of the light-strip irradiance $y(\varphi,z)$ is phase-modulated by the solid $y(\varphi,z) = \sin(\varphi_0)\rho(\varphi,z)$ for each rotation-step $\Delta\varphi$. The experimental image in Fig. 2 is formed at the CCD-camera plane corresponding to the schematics in Fig. 1 and Fig. 3.

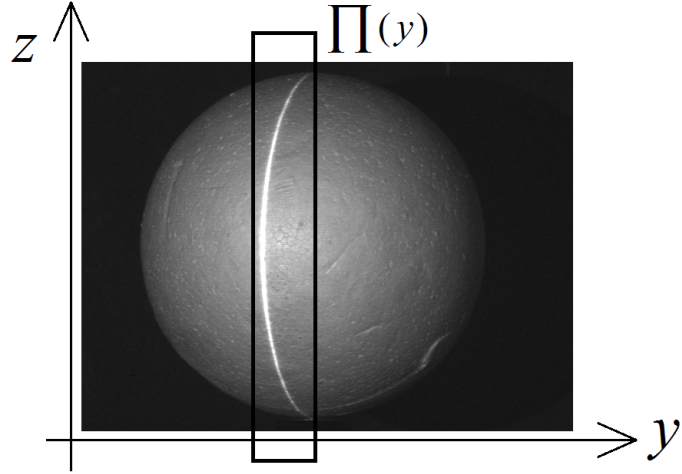

Fig. 2. A solid-sphere as imaged over the CCD-camera using a light-strip 360° profilometer. Each 360° full-digitalization, assuming a 1-degree rotation-step, needs 360 of these images. The light-strip projector is positioned as Fig. 1 shows; 360 strips (inside the square) are needed. In this figure, the ambient light is turned-on to show the digitizing sphere.

Figure 3 shows the sensitive angle $\varphi_0$ and the radius $\rho(\varphi, z_0)$ where the strip-light illuminates the solid. The distance $y(\varphi) = y(\varphi, z_0)$ is the phase-displaced strip-light as viewed from the camera. As mentioned, the mathematical relation among these functions is $y(\varphi, z) = \rho(\varphi, z)\sin(\varphi_0)$.

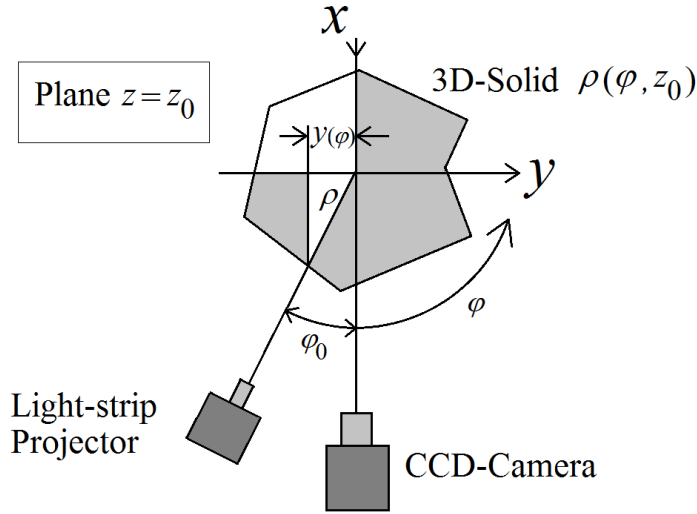

Fig. 3 This figure shows a z-cut at plane $z = z_0$ of a solid $\rho = \rho(\varphi, z)$ being digitized. The phase-displaced light-strip as seen by the camera is $\psi(\varphi) = \rho \sin(\varphi_0)$.

Once a set of $N$ individual light-strip CCD-images (see Eq. (1)) are taken at the discrete rotation angles $n\Delta\varphi$, we proceed to form a carrier-frequency fringe-pattern $I(\varphi, z)$ of the digitizing object $\rho = \rho(\varphi, z)$ in cylindrical coordinates as, (see Fig. 4),

$$I(\varphi, z) = G(\varphi, z) + G(\varphi - \Delta\varphi, z) + G(\varphi - 2\Delta\varphi, z) + \ldots + G[\varphi - (N-1)\Delta\varphi, z]; \quad \Delta\varphi = \frac{2\pi}{N}. \quad (3)$$

From Fig. 3 we see that $y = y(\varphi, z)$, so we can reduced the number of variables from $(y, z; \varphi)$ in Eq. (2) to $(\varphi, z)$ in Eq. (3). Finally Eq. (3) represents the sum of *N*-displaced light-strip irradiances constructing a carrier-frequency fringe-pattern suitable for being analyzed by the Fourier phase-demodulation technique.

As far as we know, strip-light projection profilometry coupled to the Fourier phase-demodulation is a new contribution of this paper. In the case of reference [15] by the same authors one uses a fringe-pattern projection not light-strip projection. This makes that in [15], one needs 4 phase-shifted fringe-patterns images (4-full object rotations [15]) to demodulate the phase. This is because the fringe-patterns generated in Ref. [15] have no carrier. In contrast, in this paper, only one solid revolution generates the only carrier-frequency fringe-pattern needed.

## 3. Experimental carrier-fringes construction for a centered sphere covering 360-degrees

Here we show how to construct a 360-degrees carrier-frequency fringe-pattern from *N*-Gaussian images taken within the azimuthal range of $\varphi \in [0, 2\pi)$ (a full revolution).

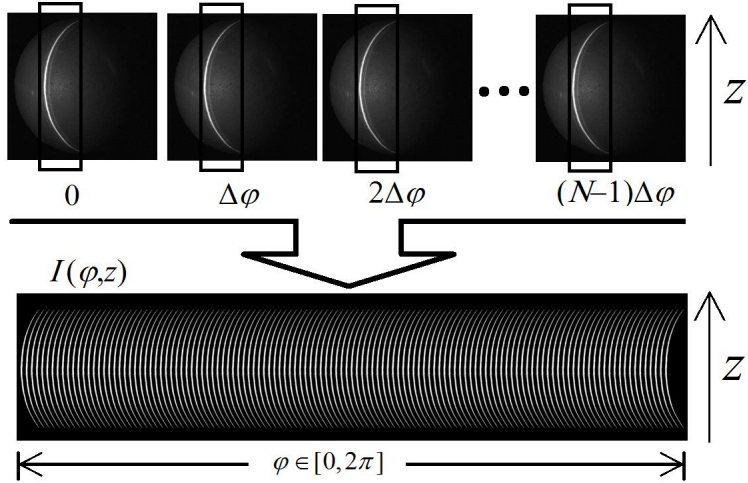

Fig. 4. 360-degrees ($\varphi \in [0, 2\pi]$) fringe-pattern construction from *N* digitized Gaussian light-strip images for each rotation step $\{0, \Delta\varphi, 2\Delta\varphi, \ldots, (N-1)\Delta\varphi\}$. Only one fringe-pattern image constructed in this way is needed to obtain the full 360-degrees digitized sphere.

Fig. 4 shows the carrier-fringe construction process followed by our light-strip 360° profilometer. As can be seen, the carrier-frequency fringe-pattern is generated by assembling *N*–Gaussian strip-lights side by side to form the carrier-fringes shown. The advantage of this scheme is that a single fringe-pattern is enough to demodulate the solid's phase. This means that we do not need to take several phase-shifted fringe-patterns as we did in Ref. [15]. As mentioned, the sphere is completely covered by the strip-light without self-occluding shadows (as it would occur in Ref. [2]). In Fig. 4, all the light-strip CCD-images over the sphere look identical because the sphere is symmetric over a full revolution (360°) around the vertical line crossing its center.

## 4. Fourier phase-demodulation of the carrier-fringes of the proposed 360° profilometer

The function space *S* of the solid surfaces that can be digitized by light-strip profilometers is given by the following set of single-valued, real functions $\rho = \rho(z, \varphi)$ as,

$$S = \left\{ \rho = \rho(\varphi, z) \mid \rho \in [0, R], z \in [-L, L], \varphi \in [0, 2\pi] \right\}, \quad \rho = \sqrt{x^2 + y^2}. \tag{4}$$

The set $S$ is bounded to $[-L, L]$ in the $z$-direction, to $[0, 2\pi]$ in the azimuthal $\varphi$ direction, and to $[0, R]$ in the $\rho$ direction (Fig.1, Fig. 3). As seen in Fig. 2, the first step in our technique is to collect $N$ images of the light-strip as imaged over the CCD-camera (Fig. 1, Fig. 2 and Fig. 3). In Fig. 2 the ambient light of the laboratory is turned-on to see the sphere, however in practice (see Fig. 4), this ambient light is turned-off. That is why the phase-modulated strip-light is the only light seen in the field of view of the CCD camera.

The sum of $N$ displaced light-strip irradiance $G(\varphi - n\Delta\varphi, z)$ forms the fringe pattern $I(\varphi, z)$ in Fig. 4. The first harmonics of $I(\varphi, z)$ may be modeled by,

$$I(\varphi, z) = \sum_{n=0}^{N-1} G(\varphi - n\Delta\varphi, z) \approx a(\varphi, z) + b(\varphi, z)\cos[\omega_0 \varphi + g\,\rho(\varphi, z)]. \tag{5}$$

The demodulated phase $g\,\rho(\varphi, z)$ gives the searched 3D-surface. The spatial-carrier $\omega_0$ of the constructed fringes from $N$-light-strip images is numerically given by

$$\omega_0 = \frac{2\pi(\textit{Number of strips in the } \varphi \textit{ direction})}{\textit{Number of pixels in the } \varphi \textit{ direction}} \left(\frac{radians}{pixels}\right) \tag{6}$$

Finally the phase $g\,\rho(z, \varphi)$ in Eq. (5) is obtained using Fourier phase-demodulation [1]. Eq. (5) may be re-written using the exponential complex form of the cosine as,

$$I(\varphi, z) = a(\varphi, z) + \frac{b(\varphi, z)}{2} e^{i[\omega_0 \varphi + g\,\rho(\varphi, z)]} + \frac{b(\varphi, z)}{2} e^{-i[\omega_0 \varphi + g\,\rho(\varphi, z)]}. \tag{7}$$

Taking the Fourier transform of this signal one obtains the spectrum of the phase-modulated fringe carrier as (see Fig. 5),

$$F[I(\varphi, z)] = A(\omega_\varphi, \omega_z) + C(\omega_\varphi - \omega_0, \omega_z) + C^*(\omega_\varphi + \omega_0, \omega_z). \tag{8}$$

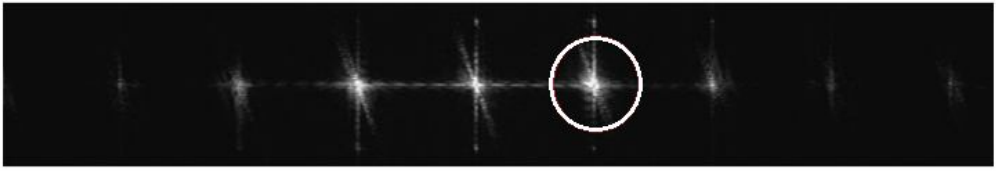

Fig. 5. This figure shows the frequency spectrum of the digitally constructed carrier-frequency fringes of the sphere shown in Fig. 4. The spectral harmonics are mainly due to the use of light-strip intensity profiles instead of a sinusoidal profile.

Where $(\omega_\varphi, \omega_z) \in [-\pi, \pi] \times [-\pi, \pi]$ is the Fourier spectrum space corresponding to the image $I(\varphi, z)$ in cylindrical coordinates (Fig. 5). The spectra $A(\omega_\varphi, \omega_z) = F[a(\varphi, z)]$ and $C(\omega_\varphi, \omega_z) = F\{(1/2)b(\varphi, z)\exp[i\,g\,\rho(\varphi, z)]\}$ are the central and right-side spectral lobes respectively. As Fig. 5 shows, the harmonics of the spectrum of $I(\varphi, z)$ are well separated so they do not interfere with the desired spectral-lobe (red-circle). The amount of harmonic distortion is minimized by locating the Gaussian light-strips in such a way that their added intensity $I(\varphi, z)$ resembles most to a sinusoidal signal. In other words, if the individual light-strips were too far away the amount of distorting harmonics would be high because the added light-strips are far from approximating a sinusoidal function. On the other hand, if the light-strips were located too close, the amplitude of the desired analytic signal would decreases. So there is a compromise between separating too-much or too-little the individual light-strip images to approximate a sinusoidal carrier-frequency fringe-pattern.

After applying the band-pass filter (the circle in Fig. 5) and displacing this spectral lobe to the spectral origin, one uses the inverse Fourier transform to obtain the desired analytic signal as,

$$F^{-1}\left[C(\omega_\varphi,\omega_z)\right] = \frac{b(\varphi,z)}{2}e^{ig\rho(\varphi,z)}. \qquad (9)$$

Finally the wrapped demodulated phase $g\rho_W(\varphi,z)$ is recovered as,

$$g\rho_W(\varphi,z) = angle\left[\frac{b(\varphi,z)}{2}e^{ig\rho(\varphi,z)}\right]; \quad \rho \in [0,R],\ z \in [-L,L],\ \varphi \in [0,2\pi]. \qquad (10)$$

We then proceed to unwrap the demodulated phase $g\rho_W(\varphi,z)$ to obtain the continuous phase $g\rho(\varphi,z)$, still in cylindrical coordinates. The final step is the rendering of the recovered solid from cylindrical coordinates $g\rho(\varphi,z)$ into a 3D-surface in Cartesian coordinates.

## 5. Conclusions

Here we have presented a new algorithm for Fourier phase-demodulation applied to 360° profilometry using strip-light projection. This new algorithm combines standard Fourier phase-demodulation profilometry with standard strip-light 360° profilometry. The herein presented light-strip profilometer is capable of digitizing solids represented by single-valued, bounded surfaces $\rho = \rho(\varphi,z)$. As any other strip-light 360° profilometer, it can digitize 3D-surfaces minimizing the self-occluding shadows.

**Acknowledgements**

The authors would like to acknowledge the financial support from project 177044 granted by the Mexican Consejo Nacional de Ciencia y Tecnologia (CONACYT).